\documentclass[aps,nofootinbib,twocolumn,floatfix,showpacs,preprintnumbers,prd]{revtex4}
\usepackage{graphicx, epsfig, bm, amsmath}
\usepackage{color}
\usepackage{hyperref}

\begin{document}

\title{Unbound geodesics from the ergosphere and potential observability of debris from ultrahigh energy particle collisions}

\author{J. Gariel$^1$\thanks{e-mail: jerome.gariel@upmc.fr}, N. O. Santos$^{1,2}$\thanks{e-mail: n.o.santos@qmul.ac.uk} and J. Silk$^{3,4,5}$\thanks{e-mail: }\\
\small{$^1$Observatoire de Paris, Universit\'e Pierre et Marie Curie,}
\small{ LERMA, UMR 8112 CNRS , 94200 Ivry sur Seine, France}\\
\small{$^2$School of Mathematical Sciences, Queen Mary,}
\small{ University of London, London E1 4NS, UK}\\
\small{$^3$Institut d\'{}Astrophysique, UMR 7095 CNRS, Universit\'e Pierre et Marie Curie,}
\small{ 98bis Boulevard Arago, 75014 Paris, France}\\
\small{$^4$Department of Physics and Astronomy, The Johns Hopkins University,}
\small{Homewood Campus, Baltimore MD 21218, USA}\\
\small{$^5$Beecroft Institute of Particle Astrophysics and Cosmology,
Department of Physics, University of Oxford, Oxford OX1 3RH, UK}}


\begin{abstract}

Particle collisions in black hole ergoregions may result in extremely high  center of mass energies that  could probe new physics if escape to infinity were possible. Here we show that some  geodesics at high inclinations to the equatorial plane may be unbound. Hence a finite flux of annihilation debris is able to escape, especially in the case of near-extremal Kerr black holes and if the Penrose process plays a role. {For a class of Penrose processes, we show  that  the Wald inequalities are satisfied, allowing the Penrose process to have a key role in high energy ejection.} Hence the possibility of observing new physics effects from a black hole accelerator  at unprecedentedly high particle collision energies remains a tantalizing, if futuristic,  experimental vision.

\end{abstract}

\maketitle

\section{Introduction}
It has been proposed  that one consequence of a density spike of cold dark matter (DM) around a supermassive black hole (BH) \cite{gondolo} is that particle collisions near the horizon can
{ occur, producing centre of mass (c.m.) energies significantly larger than those obtained in the case of particle collisions in the absence of a  BH} \cite{Banados}. This effect { (often  referred to as the  BSW effect)} is especially interesting in the case of {an extremal Kerr BH, where infinite energies can be realized, at least in principle \cite{Zaslavskii},}
a suggestion that still remains valid despite several controversial issues.

One is
a back-reaction argument against the acceleration  given by \cite{Berti,jacobson}
but { implicitly} disputed by \cite{Grib1,Tanatarov}.
{Another, for us the most urgent issue to be  addressed,
is that} of the  limiting energy at infinity and the amount of any escaping flux of energetic annihilation debris \cite{McWilliams, zaslavskii13}. We believe that  resolution of both of these issues is still incomplete. Specifically,
previous treatments generally {are restricted only to the BSW effect and most of the time only  focus on geodesics in the equatorial plane.}
However it has been noted that there is the possibility of unbound geodesics  with high energies from the ergoregion being preferentially directed near the rotation axis of a Kerr black hole \cite{Gariel,Pacheco,Gariel1} { and we believe that the Penrose process (PP) \cite{Penrose} has  a role to play in generating these high energies.}

Here we develop a simple model in the test particle approximation that allows us to estimate the fraction of  unbound geodesics {collimated along the $z$-axis} from within the ergoregion {(sections II to IV)}.
Our motivation is that collisions on near-horizon orbits around a Kerr black hole could contain unusual physics signatures, such as flavor violations, that may survive  and yield a finite flux at infinity, even though highly redshifted, due to Penrose boosting of the energetics of collisional debris in the ergosphere \cite{zaslavskiia}. We reconsider in section V the possibilities offered by the PP.

The efficiency of energy extraction from the BH by the PP
has been claimed to be low \cite{Piran,Bejger}, {mainly due to the strong restrictions imposed by the very general Wald inequalities \cite{Wald}.  But this does not exclude} the possibility of obtaining high energies
observable at infinity, due { for instance} to the ratio of the mass ejected
from the ergosphere to the mass falling into the BH {as suggested in} \cite{Harada}. { We shall see in an example studied in section V that this is effectively a possibility.}
Moreover, there are possible additional contributions, via
the decay of heavy DM particles \cite{Harada},  and of multiple collisions inside the
ergosphere \cite{Grib}.

If escape to infinity is possible even for a small fraction of the debris orbits, then observability of new physics at energies unattainable in any terrestrial particle accelerator becomes an intriguing option.

\section{Between the ergoregion and the hyperbolic limiting surface}

{ We place
ourselves in the plane $(\rho,z)$ of  Weyl coordinates $(\rho,z,\phi)$ \cite{Chandra,Carmeli} where
any spatial figure (as in Fig. 2 below)  can be easily completed in 3 dimensions by simple rotation
around the $z$-axis.}
We restrict ourselves to unbound geodesics that are able to lead to some
collimation, as studied in \cite{Gariel,Pacheco,Gariel1}. { We call "collimated" geodesics the geodesics satisfying the condition $\rho/z\ll 1$ when $z\rightarrow\infty$. This requires in particular that their angular momentum is null $L_z=0$ (see the leading term of equation (29), with equation (30), in \cite{Gariel}).}
Let us recall that these papers demonstrate the existence of Kerr geodesics
asymptotic to directions $\rho_{1}$ parallel to the $z$-axis, with $\rho_{1}^{2}=a^{2}+{\mathcal Q}/(E^{2}-1)$, where $a$
is the spin (angular momentum by unit of mass) of the BH of mass $M$
(we put $M=1$), $E$ the energy at infinity of the
particle and $\mathcal Q$ the Carter constant.

Besides the preceding  "perfectly" collimated geodesics, i.e. with asymptotes $\rho_1$, there is an infinity of (imperfectly) collimated geodesics defined from the set of parameters $(E,{\mathcal Q},L_z=0)$, or equivalently in this case $(E,\rho_1,L_z=0)$, even when $E$ and $\rho_1$ are fixed, depending on the initial conditions inside the ergosphere. We note that the geodesics are not linear equations. Though there is only one geodesic perfectly collimated for each value of $\rho_1$ (when $E$ is fixed), there is an infinity of (imperfectly) collimated geodesics for the same values of $E$ and $\rho_1$, converging towards the axis or diverging from it (e.g. see figures 2 and 5 in \cite{Gariel1}).

We begin by determining the coordinates of the point $A$ of
intersection of the trace in the plane $(\rho,z)$ of the ergoregion
with the characteristic hyperbolic surface (when it exists).
The trace in the plane $(\rho,z)$ of this hyperbolic surface is a limiting geodesic that bounds all
the collimated unbound geodesics defined for the chosen
parameters $(E,\rho_{1})$ and we call this the hyperbolic limiting
geodesic. Indeed, all  of these geodesics cannot cross this hyperbolic surface, so that, starting from the ergosphere, they are generated
inside the part of the ergosphere located between this hyperbolic surface  and the
ergosphere for $z>z_{0}$ where $z_{0}$ is the intersection of the hyperbolic surface and the $z$-axis.

The equation of the hyperbolic limiting surface
is given by (see equation (39) in \cite{Pacheco})
\begin{equation}
\left[1-\left(\frac{\rho_{1}}{a}\right)^{2}\right]^{-1}z^{2}-\left(\frac{\rho_{1}}{a}\right)^{-2}\rho^{2}=1-a^{2},  \label{2}
\end{equation}
with
\begin{equation}
\left( \frac{\rho _{1}}{a}\right) ^{2}=1-\mu _{i}^{2},  \label{3}
\end{equation}
where $\mu _{i}^{2}=-{\cal Q}/[a^{2}(E^{2}-1)]$ is a special
value of the variable $\mu \equiv \cos \theta$. The Carter
constant ${\mathcal Q}$ is necessarily negative (when positive, there is
no hyperbolic limiting surface).

The point $A$ of intersection of the ergosphere \cite{Gariel,Gariel1}, {defined by the equation
\begin{equation}
z^2=\left[1-a^2\left(1-\frac{\rho}{a}\right)\right]\left(1-\frac{\rho}{a}\right),\nonumber
\end{equation}
with the hyperbolic limiting surface (\ref{2}) is obtained from the equation
\begin{equation}
\rho^{2}-\left( 2a-\frac{1}{a}\right) \left( \frac{\rho _{1}}{a}\right)^{2}\rho -(1-a^{2})\left( \frac{\rho _{1}}{a}\right) ^{4}=0,  \label{4}
\end{equation}
with solutions
\begin{equation}
\rho_{A}=\frac{\rho_{1}^{2}}{a},\;\;\rho_{A}=\left( 1-\frac{1}{a^{2}}\right)\frac{\rho _{1}^{2}}{a}.  \label{5}
\end{equation}
In (\ref{5}) only the first solution is physically acceptable, since the
second is negative, giving
\begin{equation}
z_{A}^{2}=(1-a^{2}+\rho _{1}^{2})\left[ 1-\left( \frac{\rho _{1}}{a}\right)^{2}\right] ,  \label{6}
\end{equation}
which is always positive $({\mathcal Q}<0)$.

We can now determine the surface $\Sigma $, in the plane ($\rho ,z$), extended between the ergoregion  and the hyperbolic limiting surface,
which is given by $\Sigma =\Sigma_{1}-\Sigma_{2}$ where
\begin{equation}
\Sigma_{1}=\int_{0}^{\rho_{A}}z_{ergo}(\rho )d\rho ,\;\;\Sigma_{2}=\int_{0}^{\rho_{A}}z_{hyperbole}(\rho )d\rho .  \label{7}
\end{equation}
We now obtain
\begin{equation}
\Sigma_{1}=\frac{1}{16a^{2}}\left[ \sin (4v_{A})-\sin(4v_{0})-4(v_{A}-v_{0})\right] ,  \label{8}
\end{equation}
with $v_{0}=\arcsin (a)$ and $v_{A}=\arcsin (a\mu _{i})$, and
\begin{eqnarray}
\Sigma_{2}=(1-a^{2})\frac{\rho_{1}}{2a}\mu _{i}\left[{\mbox {arcsinh}}\left(\frac{\rho _{1}}{\sqrt{1-a^{2}}}\right)\right. \nonumber\\
\left.+\frac{\rho _{1}}{\sqrt{1-a^{2}}}
\left(1+\frac{\rho _{1}^{2}}{1-a^{2}}\right) ^{1/2}\right] .  \label{9}
\end{eqnarray}

\section{The angle of the cone formed by particles leaving to infinity}

Let us look at the influence of the BH spin $a$ on the angle of the
cone which limits the particles escaping to infinity.

The asymptote of the hyperbolic limiting surface is obtained for $\rho\rightarrow\infty$ and $z\rightarrow\infty$ and satisfies
$(\rho/z)^2\rightarrow(1-\mu_i^2)/\mu_i^2$ or $\tan^2\theta_i=\lim (\rho/z)^2$. Hence we see that, for ${\mathcal Q}$ negative, which
is the condition for $\rho_1$ to be associated with the
angle $\theta_i$, (\ref{3}) gives a direct link
between $\rho_1$ and the angle $\theta_i$ of the outgoing cone.

To see how this angle $\theta_i$ behaves with respect to $a$, we
need to express $\rho_1$ as a function of $a$. In order to do this we
adopt the case of "double roots", $r_1=r_2=Y$,
considered in \cite{Pacheco}. The difference is that here we are not going
to fix particular values of $a$ but leave it as a variable.

\begin{figure}
\includegraphics[width=\linewidth]{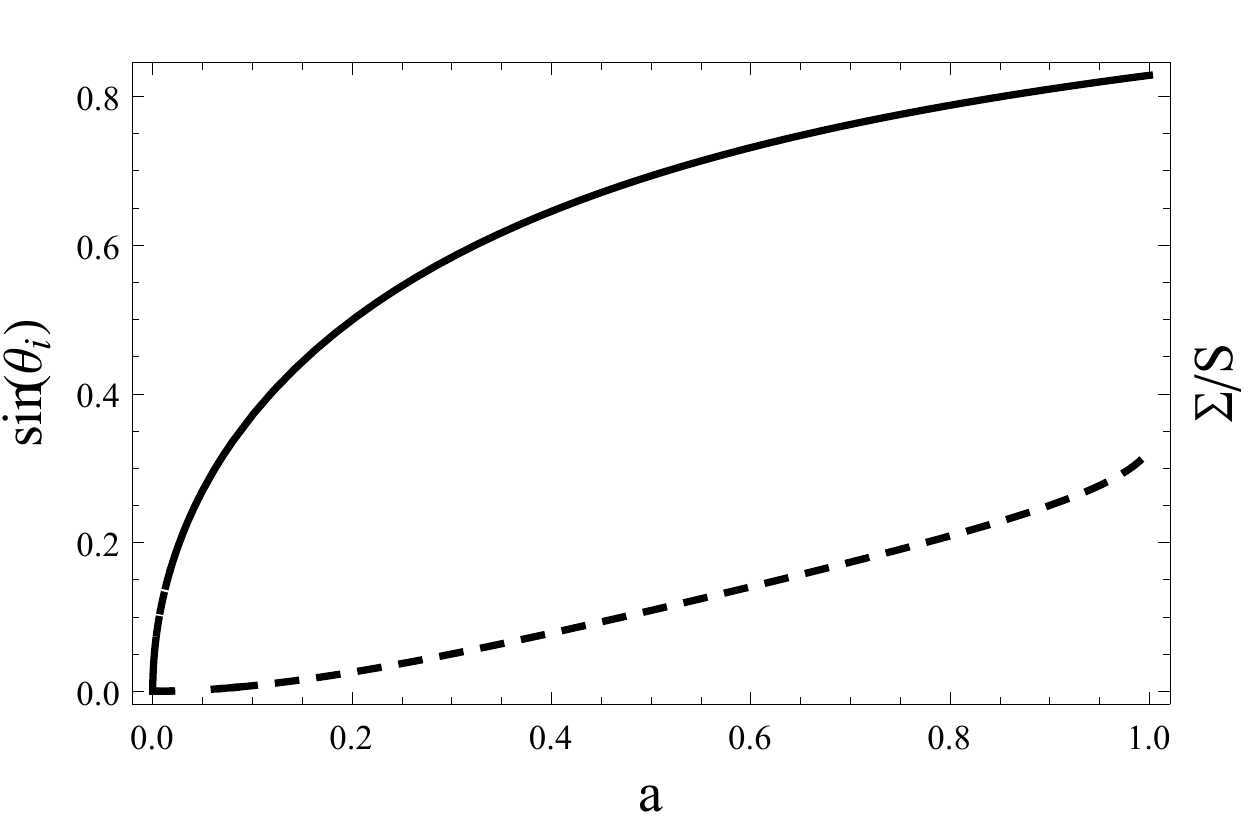}
\caption{
Left axis: plot of $\rho _{1}/a=\sin (\theta _{i})$ as a function of
the BH spin $a$, where $\theta _{i}$ is the half angle of the cone limiting
the unbound geodesics with infinite energy starting from inside the
ergosphere (solid line). Right axis:  plot of ergosphere region ratio $\Sigma (a)/S(a)$,
probability to have a particle which can in principle incur a PP and be
ejected to infinity as a function of black hole spin (dashed line).}
\label{fig:theta,spin}
\end{figure}

However we fix the energy $E$ by assuming $E\rightarrow\infty$ for two
reasons. Firstly there exists a precise value for $\rho_1$ (for a fixed $a$) for $E\rightarrow\infty$
and a small range for $\rho_1$ with $E$
rapidly decreasing. This indicates a narrow beam that is very collimated in a
domain near $\rho_1$ of particles moving to infinity and covering almost
the entire spectrum of energies \cite{Pacheco}. Secondly, the {recently found} BSW effect defined in \cite{Banados} predicts the possibility of infinite energy for particles
produced near the horizon \cite{Banados,Zaslavskii}. We note that the possibility of  a significant flux of these particles ever attaining
infinite energies is highly controversial \cite{Banados1,McWilliams}. Earlier work however has not considered detailed geodesics out of the equatorial plane. In fact, there are geodesics along or near the spin axis that are unbound.

Hence, considering $E\rightarrow \infty$ we have the following equation (13) in
\cite{Pacheco}
\begin{equation}
a^{4}(Y+1)+2a^{2}(Y-1)Y^{2}+(Y-3)Y^{4}=0.  \label{10}
\end{equation}
We find only one solution $a^{2}$ which is positive for $-1\leq Y\leq 3$,
\begin{equation}
Y^{3}-3Y^{2}+(Y+1)a^{2}=0,  \label{11}
\end{equation}
producing three solutions for $Y(a)$. If we introduce each of these
solutions into equation (14) of \cite{Pacheco} for $\rho_{1}^{2}/a^{2}$, we
obtain only one function of $a$ positive and inferior to one,
a condition equivalent to ${\cal Q}<0$. As a result, we have an expression for $\theta _{i}$, the angle of the outgoing
cone of geodesics attaining infinity, as a function of $a$: see fig. \ref{fig:theta,spin}.

\section{The admissible ergoregion}
We look for the surface $\Sigma(a)$ when $E\rightarrow\infty$
as considered previously,  i.e. via the expression $\rho_{1}(a)$
found in this case. We plot the ratio $\Sigma (a)/S(a)$ in
Fig. \ref{fig:theta,spin},
where the total surface of the ergosphere region $S(a)$ is given
by  equation (3) of \cite{Gariel1}. This ratio can be interpreted as the
probability of having a particle which can incur a PP and be
ejected to infinity with a great energy inside the limiting cone
among all the particles falling into the ergoregion.
We can see that for $a>0.8$ this probability becomes $>0.2$
and reaches about $0.33$ for $a=1$.

Consider the following numerical example. For $a=0.997$, and $E\rightarrow\infty$
which, for a double root, corresponds to $Y=-0.413334$, we
obtain $\rho_1=0.825428$, $\sin(\theta_i)=0.827911$ or
$\theta_i=0.975373\;{\mbox{rad}}=55.848^{\circ}$
and the ratio $(\Sigma /S)=0.319328$.
\begin{figure}
\includegraphics[width=\linewidth]{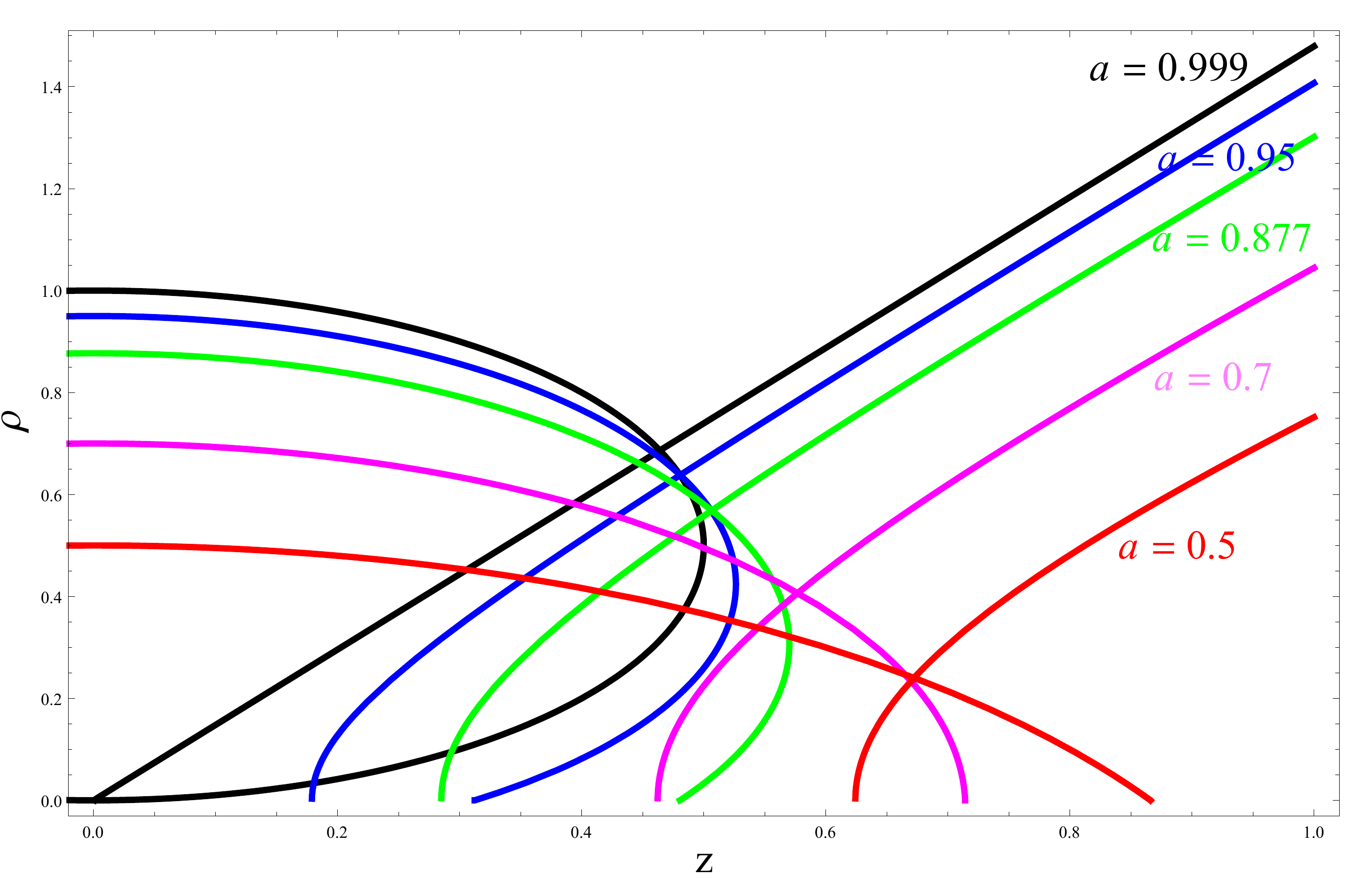}
\caption{
Plot of ergoregion and the hyperbolic  limit for the \\
cases $a$=0.5, red; 0.7, magenta; 0.877, green;
0.95, blue;  and \\
0.999, black. The admissible ergoregion for the initial conditions of particles
able to go to infinity inside the cone with a large energy is the region { of intersection}
situated between any two curves of the same color. Note that
$a=0.877$ corresponds to the maximal cross section $S$ of the
ergovolume.
The coordinates ($\rho $, $z$) are dimensionless.
The "true" corresponding distances are $\overset{\_}{\rho }=\rho M$ and $%
\overset{\_}{z}=zM$ \ where the BH\ mass $M$ is measured in number of solar
masses $M_{sol}$, itself gravitationally equivalent to the distance: $%
M_{sol}\simeq 1.48\rm km$.}
\label{fig:ergosphere}
\end{figure}

We plot in Fig. \ref{fig:ergosphere},
the ergoregion and the hyperbolic limiting surface  for  several values of $a$. See Table 1 for color code and corresponding values of $\rho_1$ and $\Sigma/S$. The values of $\rho _{1}$ corresponds to $E\rightarrow \infty $.
$\rho _{1}$ is here always $<a$ ( i.e. the Carter constant ${\cal Q}<0$).
The admissible ergosphere region for the initial conditions of particles
able to go to infinity inside the cone with a large energy is the region
situated between these two surfaces. We infer that a significant volume of the ergoregion is potentially available for launching particles to infinity.

\begin{table}
\begin{center}
\begin{tabular}{lr}
\hline\hline
$a$ & \ \  $\rho _{1}$  \ \ \ $\frac{\Sigma }{S}$ \\
\hline
$0.5$ (Red) & $0.3466$ $0.11$ \\
$0.7$ (Magenta) & $0.5336$ $0.17$ \\
$0.877$ (Green) & $0.7062$ $0.24$ \\
$0.95$ (Blue) & $0.7785$ $0.28$ \\
$0.999$ (Black) & $0.8284$ $0.33$\\
\hline\hline
\end{tabular}
\end{center}
\caption{Spin and hyperbolic limiting surface parameters (Fig. 2)}
\label{tbl:ska}
\end{table}

\section{Penrose effect and Wald inequalities}
It now remains {for us} to determine the mechanism by which these {high energy} geodesics can be populated.
{The Penrose mechanism (PP) was suggested as a mechanism for fulfilling} this role
shortly after its discovery \cite{Penrose}. {However this idea was abandoned once it was
shown how the Wald inequalities \cite{Wald}, or their generalization to collisions \cite{Bardeen}, constrained this process. The modest Penrose mechanism efficiency \cite{Piran} at  best
permits a gain of the same order of magnitude as the energy of the incoming particle \cite{Bejger,Harada1}. Thus the expected benefit of the
Penrose effect appears rather low when compared to {any similar injection} phenomenon
(particle} decay or collision) which would even occur in the absence of a Kerr BH \cite{Chandra,Wald}.

The idea of associating an external field to enlarge {the range of} the Wald  inequalities and to
increase the PP efficiency was then suggested, with, as the most likely example,
an electromagnetic field \cite{Wagh}, or magnetohydrodynamic
effects \cite{Punsly}. The Penrose-type { magnetically driven jet proposed by
Blandford and Znajek is now the most commonly accepted Kerr BH acceleration mechanism \cite{Lasota,Blandford}.}

{More recently, another effect, the so-called BSW effect was discovered \cite{Banados}. At first sight, this seems to be a good candidate for fulfilling the role of an accelerator. This effect predicts the appearance of a large local energy (in c.m.) during a collision that would occur under specific conditions}
(extremal {Kerr} BH, very near  to the horizon, in the
equatorial plane, with a precise angular momentum { derived from one of the incoming particles}).
 It has since been generalized in various forms, including to the
ergosphere outside the equatorial plane \cite{Harada} or to multiple collisions \cite{Grib}
or to the case of particles with  large angular momenta \cite{Grib1,Grib2}. {These applications are almost always in the ergosphere,} showing
that { the ergosphere} plays an important (at least local) role as "accelerator" of
particles (i.e., able to provide a high energy) \cite{Zaslavskii1}.
{ However} it has been shown that the energy provided by the latter effect, { being} local,
undergoes, because of its proximity to the centre, {a large redshift making it unobservable \cite{McWilliams,Bejger}.}

We propose a new solution to this problem, combining the BSW effect and the
PP. The remarkable feature of the PP that we put
forward is its ability to retransmit  energy to the { locally} produced high
energy { particles. The PP manages to overcome the redshift.} We will show this {via} an example.

The tension between the PP
and the Wald inequalities mainly
comes from the fact that the (true) energy $E$ of the incoming particle is
assumed { to be} positive. In the case where {the particle} comes {from infinity} at rest ($E=m$), the inequalities give a minimal speed of the falling particle
greater than $c/\sqrt{2}$ (necessary to reach a state of negative energy)
and a maximal speed for the outgoing particle lower than $c/\sqrt{2}$,
not enough to thwart the redshift. Considering an initial particle
with a negative energy permits {reversal of} these inequalities, {allowing one to
more easily obtain} an ultrarelativistic outgoing particle and a falling
particle maintained in a state of (more) negative energy.

Consider, in the ergosphere, a Penrose decay from an initial particle, with a mass $m$ and a weakly negative (true) energy $E=-m_{out}\epsilon $ ($0<\epsilon \ll 1$), into a
particle
of mass $m_{fall}<m$ falling into a more bound
state, i.e., with a more negative (true) energy $E_{fall}=-m_{fall}\epsilon $ {and with,} as we shall see,
$m_{fall}/m_{out}=\epsilon^{\prime -1}\simeq \epsilon^{-2}\gg 1$, and a particle
{of lower} mass $m_{out}$ ejected with a (true) high positive energy $E_{out}$.

Locally, in the c.m. frame, linked to the initial particle,
the conservation of linear momentum reads:
\begin{equation}
m_{fall}\gamma_{1}v_{1}-m_{out}\gamma_{2}v_{2}=0, \label{12}
\end{equation}
where $v_{i}$ is the relative velocity of the "$i$" particle ($i=1=fall$,
$i=2=out$) in the c.m. frame, and $\gamma_{i}=(1-v_{i}^{2})^{-1/2}$ its
associated Lorentz factor. Expressing the fact that the { larger} mass $m_{fall}$
has a slight recoil compared to the small mass $m_{out}$ by the relations
\begin{equation}
v_{1}=\epsilon, \;{\mbox {and}}\; \gamma_{2}=\epsilon^{-1}, \label{13}
\end{equation}
the preceding equation becomes by noting that $v_{2}=1/\gamma_{1}=(1-\epsilon^{2})^{1/2}$,
\begin{equation}
\frac{m_{out}}{m_{fall}}=\frac{\gamma_1}{v_2}\epsilon^2=\frac{\epsilon^2}{1-\epsilon^2}\simeq\epsilon^2+\epsilon^4. \label{14a}
\end{equation}
In all series expansions, we will limit ourselves to order $\epsilon^4$.

We also have
\begin{equation}
m=m_{fall}+m_{out}-\left\vert \Delta m\right\vert\stackrel{<}{\sim}m_{fall}+m_{out}, \label{15a}
\end{equation}
where the binding energy $\left\vert \Delta m\right\vert$ is assumed to be
very small compared to the smallest mass , i.e., $\left\vert\Delta m\right\vert \ll m_{out}\ll m_{fall}$ or $m$. For example, it is sufficient to take
\begin{equation}
\frac{\left\vert\Delta m\right\vert}{m_{out}}=\epsilon. \label{16a}
\end{equation}
Hence, substituting (\ref{16a}) into (\ref{15a}) gives
\begin{equation}
\frac{m_{fall}}{m}=1-\frac{m_{out}}{m}+\frac{m_{out}}{m}\epsilon. \label{17a}
\end{equation}
On the other hand, using (\ref{14a}) and (\ref{17a}), we have
\begin{equation}
\frac{m_{out}}{m}=\frac{m_{out}}{m_{fall}}\frac{m_{fall}}{m}=\frac{\epsilon ^{2}}{1-\epsilon ^{2}}\left[1-\frac{m_{out}}{m}(1-\epsilon)\right], \label{18a}
\end{equation}
or
\begin{equation}
\frac{m_{out}}{m}\left[1+\frac{\epsilon^2(1-\epsilon)}{1-\epsilon^2}\right]=\frac{\epsilon^2}{1-\epsilon^2}, \label{19a}
\end{equation}
i.e.,
\begin{equation}
\frac{m_{out}}{m}=\frac{\epsilon^2}{1-\epsilon^3}\simeq\epsilon^2. \label{20a}
\end{equation}
From (\ref{17a}) we also obtain
\begin{equation}
\frac{m_{fall}}{m}=\frac{1-\epsilon^2}{1-\epsilon^3}\simeq1-\epsilon^2+\epsilon^3. \label{21a}
\end{equation}

Secondly, we have the Penrose equality
\begin{equation}
E_{out}=\left\vert E_{fall}\right\vert -\left\vert E\right\vert, \label{22}
\end{equation}
and to have a significant Penrose effect, the condition $\left\vert
E_{fall}\right\vert \gg \left\vert E\right\vert $ is necessary, consistently
with what we saw earlier, $m_{out}/m_{fall}\simeq \epsilon^{2}+\epsilon ^{4}$. Thus, in this example, the mass effect is the source
of the Penrose effect. The Penrose equality becomes
\begin{eqnarray}
\Gamma_{out} &\equiv &\frac{E_{out}}{m_{out}}=\frac{\left\vert
E_{fall}\right\vert }{m_{fall}}\frac{m_{fall}}{m_{out}}-\frac{\left\vert
E\right\vert }{m_{out}} \nonumber \\
&=&\frac{1}{\epsilon }-2\epsilon. \label{23}
\end{eqnarray}
Finally, consider the Wald inequalities (equation (4) in \cite{Wald}).
For the particle $m_{fall}$ we have
\begin{eqnarray}
-\gamma_{1}\frac{m_{out}}{m}\epsilon-\gamma_{1}v_{1}\left(\frac{E^{2}}{m_{out}^{2}}\frac{m_{out}^{2}}{m^{2}}+1\right)^{1/2}
\leq-\epsilon \nonumber\\
\leq
-\gamma_{1}\frac{m_{out}}{m}\epsilon+\gamma_{1}v_{1}(\frac{E^{2}}{m_{out}^{2}}\frac{m_{out}^{2}}{m^{2}}+1)^{1/2}, \label{24}\\
\gamma_{1}\epsilon^{3}+\gamma_{1}\epsilon(\epsilon^{6}+1)^{1/2}\geq \epsilon\geq\gamma_{1}\epsilon^{3}
-\gamma_{1}\epsilon (\epsilon^{6}+1)^{1/2}, \label{25}\\
\gamma_{1}\epsilon^{2}-\gamma_{1}\left(1+\frac{\epsilon^{6}}{2}\right)\leq
1\leq\gamma_{1}\epsilon^{2}+\gamma_{1}\left(1+\frac{\epsilon^{6}}{2}\right), \label{26}\\
-\gamma_{1}\leq 1\leq\gamma_{1}, \label{27}
\end{eqnarray}
which is always verified for any $v_{1}$. While for the particle $m_{out}$ we have
\begin{eqnarray}
-\gamma_{2}\frac{m_{out}}{m}\epsilon-\gamma_{2}v_{2}(1+\epsilon^{6})^{1/2}\leq\frac{E_{out}}{m_{out}} \nonumber\\
\leq-\gamma_{2}\frac{m_{out}}{m}\epsilon+\gamma_{2}v_{2}(1+\epsilon^{6})^{1/2}. \label{27}
\end{eqnarray}
The first inequality of (\ref{27}) is always verified (negative left, positive right), and
the second one is
\begin{equation}
\Gamma_{out}\leq\gamma_{2}v_{2}-\gamma _{2}\epsilon^{3}+\gamma _{2}\frac{\epsilon^{6}}{2}, \label{28}
\end{equation}
or
\begin{equation}
\Gamma _{out}=\frac{1}{\epsilon }-2\epsilon <\frac{1}{\epsilon}\left(1-\frac{\epsilon^{2}}{2}\right)-\epsilon^{2}
=\frac{1}{\epsilon }-\frac{\epsilon}{2}-\epsilon^{2}, \label{29}
\end{equation}
or $2\epsilon >\epsilon/2$, which is always true. Therefore,
the Penrose equality agrees with the Wald inequalities, and we can take
\begin{equation}
\Gamma_{out}\simeq \frac{1}{\epsilon}=\gamma_{2}.
\end{equation}

Thus we see that for large values of $\gamma_{2}$, or equivalently
values $v_{2}$ approaching $1$, due to a mass effect, the particle "out"
can have a high energy at infinity (which can be a bit lower than that
obtained locally $m_{out}\gamma _{2}$), in agreement with the Penrose equality
and unrestricted by the Wald inequalities.

{As an immediate consequence,} if we define the efficiency of the PP as $\eta^{\prime }=E_{out}/\left\vert E\right\vert $, we have
$\eta^{\prime}=m_{out}\Gamma _{out}/m_{out}\epsilon =1/\epsilon ^{2}-2\simeq 1/\epsilon ^{2}$, i.e., we obtain a large efficiency.

 Let us stress the remarkable result} that the same calculations made from the hypothesis of {an initial particle}
with a slightly positive energy $E=+m_{out}\epsilon$
would lead to violated Wald inequalities. The limiting case $E=0$ respects the inequalities.

The questions that remain { concerning the initial particle} are: (i) How
{ was such an initial particle inside the ergosphere able to reach a state of negative energy}
(in our example above,
slightly negative: $\left\vert E\right\vert \ll m_{out}$)? and (ii) is it possible in a decay to have high speed ejection?

It seems to us that {we can} obtain the first result from
effects of BSW-type. They are purely gravitational and occur only inside the
ergosphere. Though directly unobservable because of the redshift \cite{McWilliams}, the BSW effect { locally produces (in the c.m. frame)} high energy {and} can afford
to send a particle, for example coming from infinity at rest ($E=m$), at a
relative speed greater than $1/\surd 2=0.7$ sufficient to enter a state of
negative energy \cite{Wald}. Thus, the BSW effect, or those of { similar type},
can be understood as possible "triggers" of the PP. More
precisely, they are able to implement the necessary condition (state of
negative energy) on a particle they produce to trigger a PP.

{Concerning} the second question, {we note} that local physical phenomena can happen,
such as  $\beta^-$-decays, which can eject particles {of} small mass
(electron, antineutrino), { relative to the mass of the expelling particle (radioactive nuclei),}
with ultrarelativistic speed $v_{2}>0.9$, and a weak
effect of recoil ($v_{1}\ll 1$). The recoil is sufficient to maintain the
large particle in a more "connected" (with regard to the BH) state, that is to say with
negative energy higher (in absolute value) than that of the initial
particle, triggering the PP. Another example could be the annihilation of DM. For {a} hypothetical DM particle of mass 100GeV (see for example \cite{Gorchtein}), the Lorentz factor of a produced outgoing electron can be as high as $\sim 10^3$.

It is likely that Penrose injection will not uniformly fill the ergoregion but may be preferentially enhanced towards the equatorial plane as $a$ increases { \cite{Banados1}}. This would reduce the effect we  consider, since the effective volume available for unbound orbits would be reduced.

\section{Conclusion}
Our calculations demonstrate that the intersection of a cone along the spin axis, described by us as the hyperbolic limiting surface,  with the ergoregion allows the launching of collimated unbound geodesics with high energies. { The { "two-step"} process (BSW effect triggering PP) that we propose shows that it is possible to populate such geodesics.} This would boost any potential signal at infinity.
Suppose that the ergoregion is  populated with particles. Then we can use the product of the ratio of the intersecting volume to the ergoregion with the total annihilating flux in the ergosphere as a crude measure of the flux generated by particle collisions. We conclude that a finite flux of annihilation debris is able to escape to infinity, in the case of near-extremal Kerr black holes.
Hence the possibility of observing new physics effects from a black hole accelerator  at unprecedentedly high particle collision energies remains a tantalizing if futuristic
experimental vision.

\section*{Acknowledgments}
One of us (JS) acknowledges useful discussions with T. Jacobson and W. Unruh.
The research of JS has been supported at IAP by  the ERC project  267117 (DARK)
hosted by Universit\'e Pierre et Marie Curie - Paris 6   and at JHU by NSF grant OIA-1124403.

\end{document}